\documentclass[twocolumn,prb,aps,amsmath,amssymb]{revtex4}

\usepackage{graphicx}
\usepackage{dcolumn}
\usepackage{bm}

\begin{document}

\preprint{Applied Physics Letter}

\title{Signal processing by opto-optical interactions between self-localized and free propagating beams in liquid crystals}

\author{Alessia Pasquazi, Alessandro Alberucci, Marco Peccianti and Gaetano Assanto}

\affiliation{
NooEL - Nonlinear Optics and OptoElectronics Laboratory 
\homepage{http://optow.ele.uniroma3.it}
Department of Electronic Engineering and National Institute for the Physics of Matter - INFM-CNISM 
University ``Roma Tre'', Via della Vasca Navale 84, 00146 Rome - Italy}

\date{\today}

\pacs{42.65.Tg, 42.65.Jx, 42.70.Df,}
\begin{abstract}
The reorientational nonlinearity of nematic liquid crystals enables a self-localized spatial soliton and its waveguide to be deflected or destroyed by a control beam propagating across the cell. We demonstrate a simple all-optical readdressing scheme by exploiting the lens-like perturbation induced by an external beam on both a nematicon and a co-polarized guided signal of different wavelength. Angular steering as large as 2.2 degrees was obtained for control powers as low as 32mW in the near infrared.
\end{abstract}

\maketitle

Optical spatial solitons, i.e. self-localized light beams in nonlinear media, are excellent building blocks for all-optical signal processing. In Kerr-like media with a self-focusing response, spatial solitons can also confine a signal which propagates un-diffracted in the corresponding waveguide. \cite{trillo, stegeman286-1518-1999,Kivshar,conti} Owing to their ``particle-like behavior'', spatial solitons have been exploited in several configurations and materials for optical signal readdressing, logic gating and switching. \cite{trillo,shalaby16-1472-1991,torruellas68-1449-1996,kang21-189-1996,stegeman286-1518-1999,Kivshar,shih21-1538-1996,shih78-2551-1997,krolikowski10-823-1998,aitchinson18-1153-1993}

In most configurations, however, spatial solitons were required to interact with other solitons or beams over propagation lengths orders of magnitude larger than their transverse size. 
Recently, the highly non local and non-resonant molecular response of undoped nematic liquid crystals (NLC) has enabled the demonstration of stable (2+1)-dimensional spatial solitons (or {\it nematicon} \cite{assanto14-44-2003}) and their interactions at mW power levels.\cite{assanto14-44-2003,peccianti26-1690-2001,peccianti81-3335-2002,conti92-113902-2004,peccianti432-733-2004}
In this Letter, based on the large molecular nonlinearity of NLC (several orders of magnitude higher than in CS$_2$),\cite{khoo,tabyrian136-1-1986} we demonstrate all-optical readdressing of spatial solitons via the lens-like perturbation induced by an external beam propagating across the medium.\par
NLC consist of anisotropic rod-like molecules which tend to be aligned in a specific direction owing to intermolecular forces, anchoring at the interfaces, an electrostatic (or in general low-frequency) or magnetostatic bias. \cite{gennes}  Fig. 1(a) shows a typical cell arrangement in the presence of an external voltage to pre-tilt the molecules and facilitate their all-optical response.\cite{peccianti26-1690-2001} An extraordinary-polarized light beam can reorient the molecules towards the field vector due to (induced) dipolar reaction. Such reorientation results in a self-focusing nonlinearity (an extraordinary index increase) because of the medium birefringence; this response is also non local because of the elastic forces binding the molecules to one another. \cite{conti92-113902-2004} When diffraction is balanced by nonlinearity, a {\it nematicon} is obtained: its own graded-index waveguide also able to guide a co-polarized (weak) signal even at a different wavelength (see Fig. 1(b)).\cite{assanto14-44-2003} 
\begin{figure}
\includegraphics[width=8.5cm]{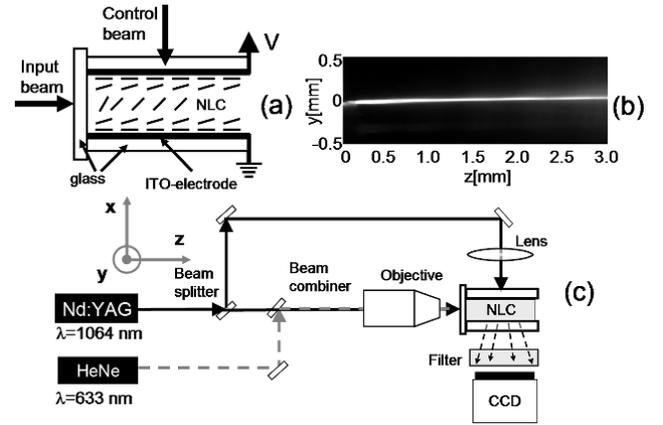}
\caption{(a) Geometry of the planar cell under voltage bias; (b) NLC self-focusing response in the absence of a control beam: an x-polarized weak signal (35$\mu$W at 633nm) is confined in the {\it nematicon} generated by a 2.2mW near-infrared beam (1064nm). Here the external voltage was $V$=1.4V across a 75$\mu$m thick cell. (c) Experimental setup} 
\label{setup}
\end{figure}

An external \textit{z}-polarized (control) beam propagating through the NLC cell can also induce an index perturbation owing to reorientation (see sketch in Fig. 1 (a) and (c)). 
Such perturbation can overlap the refractive distribution due to the voltage bias and the {\it nematicon}, when launched.
Defining $\theta$ the angle between the light wave vector and the molecular major axis (or {\it director}), the perturbation can be evaluated by considering both the low-frequency bias and the control beam of electric field amplitudes E$_{RF}$ and E$_{c}$, respectively:
\begin{equation}
K\nabla^2\theta+\epsilon_0(\frac{1}{2}\Delta\epsilon_{RF}|E_{RF}|^2-\frac{1}{4}\epsilon_a |E_{c}|^2 )\sin(2\theta)=0
\label{reorientation_eq}
\end{equation}
as well as the propagation of the external beam across the non-homogeneous sample: \cite{khoo,tabyrian136-1-1986}

\begin{equation}
2ik\frac{\partial E_c}{\partial x}+\nabla^2_{yz}E_c+k_0^2(n(\theta)^2-\bar{n}^2 )E_c =0
\label{prop_eq}
\end{equation}

being $\Delta\epsilon_{RF}$ and $\epsilon_a$ the dielectric anisotropies at low and optical frequencies, respectively, {\it K} the NLC elastic constant (averaged over all molecular distortions),\cite{gennes,khoo} \textit{k}=2$\pi\bar{n}/\lambda$=\textit{k}$_0\bar{n}$ the wave number with $\lambda$ the wavelength, $\bar n$ and n($\theta$) a reference and the extraordinary refractive indices, respectively. 
Coupled equations \ref{reorientation_eq} and \ref{prop_eq} were numerically integrated with reference to a cell of thickness \textit{h}=75$\mu$m, the liquid crystal E7 (\textit{K}=1.9$\times$10$^{-11}$N, $\Delta\epsilon_{RF}$=14.5, $\epsilon_a$=0.61) and $\lambda$=1.064$\mu$m. The boundary conditions corresponding to our experimental arrangement were $\theta$(0,\textit{y},\textit{z})=$\theta$(\textit{h},\textit{y},\textit{z})=2$^\circ$, a bias \textit{V}=1.4V, a Gaussian control beam of given waist \textit{w}$_c$ and power P$_c$.
\begin{figure}
\includegraphics[width=8.5cm]{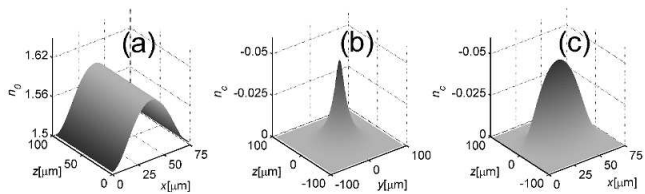}
\caption{Computed refractive index in the NLC cell.  (a) Index distribution due to an applied voltage \textit{V}=1.4V. Nonlinear perturbation due to a control beam (\textit{P}$_c$=32mW and \textit{w}$_c$=8$\mu$m) across (b) the mid-plane (\textit{h}/2,\textit{y},\textit{z}) and (c) the transverse plane (\textit{x},0,\textit{z}), respectively. In \textit{x}=0 and \textit{x}=h the anchoring forces the index perturbation to zero.}
\label{lens}
\end{figure}
Figure \ref{lens}(a) shows the refractive distribution due to the external voltage \textit{V} and experienced by a \textit{z}-propagating wave polarized in the \textit{xz} (principal) plane, i.e. corresponding to or guided by a {\it nematicon}). Figs. \ref{lens}(b-c) display the effect of a control beam (of waist 8$\mu$m and power 32mW) which, owing to its polarization, tends to reduce $\theta$ and determine a diverging (defocusing) lens-like perturbation with a negative index peak of 0.049.  When a spatial soliton is excited, therefore, this negative lens close to or on its path is expected to interfere with its propagation, destroying or deviating it according to their relative position.  The same considerations apply to the co-polarized signal eventually guided by the {\it nematicon}.\par
In the experiments we employed a sample with the nematic E7 and the parameters specified above. The planar cell, realized in glass (see Fig. 1(a)), was equipped with Indium-Tin-Oxide (ITO) electrodes for the application of a bias. The setup consisted of a Nd:YAG pump collinear with a He-Ne probe and injected into the cell (along \textit{z}) in the extraordinary polarization to excite a {\it nematicon}. A $\lambda$=1064nm \textit{z}-polarized control beam was focused in the NLC as it propagated across the cell thickness (along \textit{x}). The control spot could be precisely positioned with respect to the soliton trajectory in the \textit{yz} plane.
In the absence of the control beam, typical results for a 35$\mu$W probe co-launched and co-polarized with the 1.064$\mu$m pump are displayed in Fig. 1(b). When a {\textit{nematicon}} was generated for input powers as low as 2.2mW and \textit{V}=1.4V, the weak signal propagated undiffracted as a guided mode on the nonlinear waveguide for propagation distances well exceeding the Rayleigh length.\cite{assanto14-44-2003}

\begin{figure}
\includegraphics[width=8.5cm]{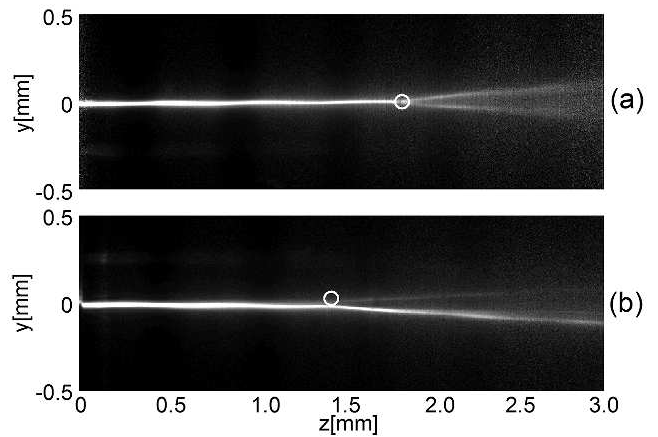}
\caption{Effect of the control beam of power \textit{P}$_c$=32mW on the signal guided by a {\textit{nematicon}} in the \textit{yz} plane. (a) When the nonlinear lens (its position indicated by the circle) overlapped with the soliton, the confined beam split up into a Y-junction, its guiding ability after the stem depending on the soliton power. (b) As the lens was shifted upwards by $\Delta$\textit{y}=9$\mu$m the soliton and the signal were steered downwards.}
\label{split} 
\end{figure}

When the control beam was turned on and focused nearby the {\it nematicon} we could observe two main phenomena, as shown in Fig. \ref{split}. When the nonlinear lens was formed on the soliton trajectory, the latter split up into two symmetrically displaced arms, thereby forming a Y junction. The guided He-Ne probe underwent the same behavior, as visible in Fig. \ref{split}(a). Noticeably, the two beams after the Y-stem were self-localized or diffracting depending on the amount of power associated to the {\textit{ nematicon}}, in substantial agreement with the behavior predicted for the interaction of spatial solitons with localized inhomogeneities.\cite{burtsev52-4474-1995}  As the perturbing lens was shifted along \textit{y}, the soliton and its signal were steered towards the higher index region, bending in the half plane (\textit{yz}) not containing  the external beam. Fig. \ref{split}(b) shows such a case corresponding to \textit{P}$_c$=32mW and a \textit{y}-shift of 9$\mu$m.  In our geometry the latter value maximized
d the sought steering for powers never exceeding 32mW, i.e. low enough not to induce thermal effects. We could ascertain the negative sign of the induced nonlinear index change based on standard analysis of the diffraction ring-like patterns induced by a light beam traversing an NLC layer,\cite{khoo,santamato9-564-1984} whereas its reorientational nature was confirmed by time-resolved pump-probe experiments, the turn-on time being much faster for a thermally-induced response.\\
Fig. 4 graphs various {\it nematicon} trajectories as the control power was varied. In each curve, noise and slight oscillations are due to thermal fluctuations of the medium as well as dis-uniformities in the cell. Finally, Fig. \ref{anglevspower} plots the steering angle (with respect to the unperturbed case) versus \textit{P}$_c$: a maximum deflection of 2.2$^\circ$ was achieved for \textit{P}$_c$=32mW. \par

\begin{figure}
\includegraphics[width=7.5cm]{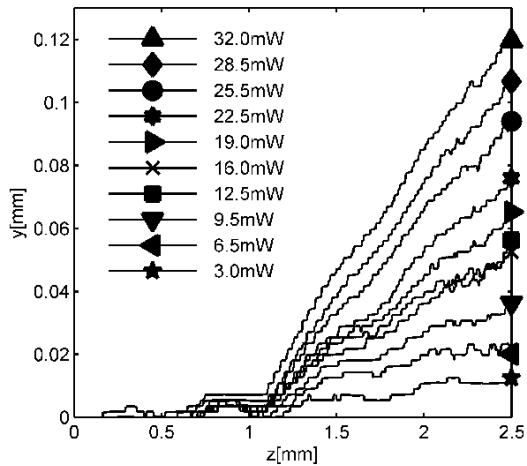}
\caption{{\it Nematicon} trajectories in the \textit{yz} plane for various \textit{P}$_c$. The control beam had a waist of 8 $\mu$m and was centered in $\Delta$\textit{y}=9$\mu$m.}
\label{graphs} 
\end{figure}

In conclusion, we have demonstrated all-optical routing based on the nonlinear interaction between orthogonally propagating beams in nematic liquid crystals. A {\textit{ nematicon}} and the signal it confined could be deviated using the negative lens-like perturbation induced by a control beam. The externally induced GRIN did not appreciably affect the guiding properties of the soliton, solely intervening on its trajectory except for the head-on geometry (Fig. \ref{split}(a)).\par
Noteworthy, compared to previous soliton steering, the demonstrated scheme requires a very short interaction region. Thereby repeated all-optical deflections and more complex logic and routing schemes could be performed based on the interaction between spatial solitons and multiple independent perturbations analogous to the one exploited in this work. \cite{assanto14-44-2003} \\
\newpage

Acknowledgments \\
GA thanks N. Tabiryan (Beam Engineering-USA) for enlightening discussions.

\begin{figure}
\includegraphics[width=7cm]{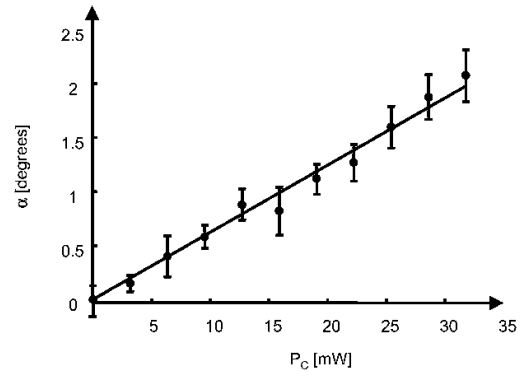}
\caption{Steering angle $\alpha$ versus control power \textit{P}$_c$. Parameters are as in Fig. \ref{graphs}.}
\label{anglevspower}
\end{figure}

\bibliography{apssamp}

\end{document}